\documentclass[usenatbib,iop,numberedappendix]{aeb_emulateapj_2010}

\usepackage[usenames,dvipsnames]{color}
\usepackage{amsmath}
\usepackage{amssymb}
\usepackage{xspace}
\usepackage[normalem]{ulem}

\usepackage{natbib}

\def\del#1{{}}

\newdimen\sa  \def\sd{\sa=.1em  \ifmmode $\rlap{.}$''$\kern -\sa$
                                \else \rlap{.}$''$\kern -\sa\fi}
 \def\vss{\vskip 4pt}

\def\s{{\rm s}} %...........seconds
\def\yr{{\rm yr}} %.........years
\def\erg{{\rm erg}} %.......ergs
\def\K{{\rm K}} %...........Kelvin

\def\Lj{L_{\rm jet}}
\def\etaj{\eta_{\rm jet}}
\def\ergs{{\rm erg\,s^{-1}}}
\def\Pa{P_{\rm acc}}
\def\req{r_{\rm eq}}

\begin{document}

\title{
The Event Horizon of M87
}

\author{
Avery E.~Broderick\altaffilmark{1,2},
Ramesh Narayan\altaffilmark{3},
John Kormendy\altaffilmark{4,5,6},\\
Eric S.~Perlman\altaffilmark{7},
Marcia J.~Rieke\altaffilmark{8},
and
Sheperd S.~Doeleman\altaffilmark{3,9}
}
\altaffiltext{1}{Perimeter Institute for Theoretical Physics, 31 Caroline Street North, Waterloo, ON, N2L 2Y5, Canada}
\altaffiltext{2}{Department of Physics and Astronomy, University of Waterloo, 200 University Avenue West, Waterloo, ON, N2L 3G1, Canada}
\altaffiltext{3}{Harvard-Smithsonian Center for Astrophysics, 60 Garden Street, Cambridge, MA, 02138}
\altaffiltext{4}{Department of Astronomy, University of Texas at Austin, 2515 Speedway, Mail Stop C1400, Austin, TX 78712-1205, USA; kormendy@astro.as.utexas.edu}
\altaffiltext{5}{Max-Planck-Institut f\"ur Extraterrestrische Physik, Giessenbachstrasse, D-85748 Garching-bei-M\"unchen, Germany}
\altaffiltext{6}{Universit\"ats-Sternwarte, Scheinerstrasse 1, D-81679 M\"unchen, Germany}
\altaffiltext{7}{Department of Physics and Space Sciences, 150 W. University Blvd., Florida Institute of Technology, Melbourne, FL 32901, USA; eperlman@fit.edu}
\altaffiltext{8}{Steward Observatory, University of Arizona, 933 North Cherry Avenue, Tucson, AZ 85721-0065}
\altaffiltext{9}{MIT Haystack Observatory, Off Route 40, Westford, MA 01886, USA}

\shorttitle{The Event Horizon of M87}
\shortauthors{Broderick et al.}

\begin{abstract}
The $6\times10^9\,M_\odot$ supermassive black hole at the center of
the giant elliptical galaxy M87 powers a relativistic jet.
Observations at millimeter wavelengths with the Event Horizon
Telescope have localized the emission from the base of this jet to
angular scales comparable to the putative black hole horizon.  The jet
might be powered directly by an accretion disk or by electromagnetic
extraction of the rotational energy of the black hole.  However, even
the latter mechanism requires a confining thick accretion disk to
maintain the required magnetic flux near the black hole. Therefore,
regardless of the jet mechanism, the observed jet power in M87 implies
a certain minimum mass accretion rate.  If the central compact object
in M87 were not a black hole but had a surface, this accretion would
result in considerable thermal near-infrared and optical emission from
the surface.  Current flux limits on the nucleus of M87 strongly
constrain any such surface emission. This rules out the presence of a
surface and thereby provides indirect evidence for an event horizon.
\end{abstract}


\maketitle

\section{Introduction} \label{sec:I}

It is now widely accepted that active galactic nuclei (AGN) are
powered by supermassive objects (reaching $10^{10}\,M_\odot$) that are
sufficiently compact to exclude all other astrophysically credible
alternatives to black holes \citep{1984ARA&A..22..471R}.  However, it
is less clear that these objects possess the defining characteristic
of a black hole: an event horizon\footnote{Here we will employ an
  astrophysically motivated definition of the horizon: a surface from
  inside which astronomical signals cannot propagate to large
  distances in astronomically relevant timescales.  Formally, for a
  dynamical system, such a surface is identified with the apparent
  horizon. However, in the context of astrophysical black holes
  described by general relativity, it corresponds to the event horizon
  as well.}.  The existence of black hole event horizons plays a
central role in a number of puzzles associated with black holes, e.g.,
the information paradox. A number of recent results suggest that a
resolution of these puzzles may result in modifications on horizon
scales
\citep[e.g.,][]{2011CQGra..28l5010M,2013JHEP...02..062A,2014arXiv1406.0807M},
which provides strong motivation for seeking astronomical evidence for
the reality of event horizons.

Accretion onto compact objects with a surface, e.g., white dwarfs,
neutrons stars, results in the formation of a boundary layer in which
any remaining kinetic energy contained within the accretion flow is
thermalized and radiated.  In contrast, gas accreting onto a black
hole is free to advect any excess energy across the horizon without
further observational consequence.  If the mass accretion rate,
$\dot{M}$, can be independently estimated, this difference provides an
observational means to distinguish between the presence of a surface,
or more accurately a ``photosphere,'' and a horizon.

The above argument has already been used to argue for the existence of
event horizons in X-ray binaries by comparing neutron star and black
hole systems in aggregate \citep{N97,G01,NH02,Done_Gierlinski_2003,NM08}.  However, the
advent of horizon-resolving observations, enabled by
millimeter-wavelength very long baseline intererometric observations
(mm-VLBI) carried out by the Event Horizon Telescope
\citep[EHT,][]{2009astro2010S..68D,2010evn..confE..53D,Doel_etal:08,Fish_etal:11,Doel_etal:12}, has made it
possible to extended the argument to individual systems.  This is
primarily because restricting the size of any photospheric emission to
horizon scales enables two important simplifications:
\begin{enumerate}
\item Any putative photosphere that lies within the photon orbit is
  expected to radiate to a good approximation like a blackbody,
  independently of the details of its composition \citep{BN06,BN07}. This is because
  a majority of the photons emitted from the photosphere will be
  strongly lensed back onto the photosphere, thermally coupling the
  photosphere to itself and to the emitted photon field.  As the
  redshift of the surface increases, the blackbody approximation
  becomes increasingly accurate.
\item The expected temperature of the photosphere emission, as seen by
  distant observers, is dependent upon the mass accretion rate
  $\dot{M}$ and the apparent photosphere size, the latter of which is
  fixed when the photosphere lies within the photon orbit.  Thus,
  assuming that the system has reached steady state\footnote{The
    additional gravitational time delay for radiation to escape from a
    compact surface is insufficient to prevent the system from
    reaching steady state, since the timescale diverges only
    logarithmically as the radius of the surface approaches the
    horizon radius \citep{BN06,NM08,BLN09}.}, any independent estimate
  of $\dot{M}$ immediately determines both the luminosity and the
  radiation spectrum as seen by a distant observer.
\end{enumerate}
Essentially, by restricting the surface to be sufficiently compact, it
is possible to robustly predict the properties of any putative
photosphere emission, independent of the specific properties of the
radiating surface.  Direct flux limits can then be used to constrain
and/or exclude the presence of a photosphere.

The above argument has already been successfully employed in the case
of Sagittarius A* (Sgr A*), the supermassive black hole at the center
of the Milky Way \citep{Nara_etal:98,BN06,BN07,NM08,BLN09}, where
limits on $\dot{M}$ were obtained from the observed bolometric
luminosity, assumed to arise from the accretion flow during infall.
It was shown that, for physically reasonable radiative efficiencies,
it is impossible to accommodate a photosphere. Therefore, Sgr A* must
have an event horizon behind which the kinetic energy of the infalling
accreting gas is hidden.

The supermassive black hole at the center of the nearby giant
elliptical galaxy M87 (which we will refer to as M87* hereafter) is a
second target for which the EHT has provided horizon-scale limits upon
the extent of its mm-wavelength emission.  Here we explore the
implications of these and related observations for the existence of an
event horizon in M87*.  In what follows we assume a mass of
$6.16\times10^9\,M_\odot$ and distance of $16.5$~Mpc for M87*,
reported in the recent review by \citet{KH13}.  The mass estimate is
based on the stellar dynamical modeling described in
\citet{2011ApJ...729..119G}, and is roughly a factor of two larger
than the value obtained by gas dynamical measurements
\citep{2013ApJ...770...86W}; both methods are potentially complicated
by the fact that M87* is offset from the center-of-light by as much as
10~pc, impacting the underlying assumptions regarding orbital isotropy
\citep{Batc_etal:10}.  However, our qualitative conclusions are
insensitive to this difference, being marginally stronger with the
smaller black hole mass.\footnote{The temperature of a putative 
  surface observed at infinity, defined in Equation \ref{eq:Tinf},
  depends on the mass as $M^{-1/2}$, and thus is higher for smaller
  masses.  This shifts the resulting emission to higher frequencies
  where better limits on the observed nuclear emission exist.}

M87* is nearly three orders of magnitude more massive than Sgr A*,
probing a mass regime more relevant for the bright AGN observed at
high redshift.  Unlike Sgr A*, whose mm and radio emission seem to be
primarily from the accretion flow \citep{YN14}, the mm/radio emission
of M87* is dominated by a powerful relativistic jet.

Two lines of argument strongly imply that the relativistic jet in M87
originates near the black hole.  First, astrometric measurements of
the radio core position reveal a wavelength-dependent shift,
asymptoting to a fixed position at short wavelengths, consistent with
a black-hole-launched jet \citep{2011Natur.477..185H}.  Second, the
small scales implied by EHT observations at 1.3mm are commensurate
with the scales expected near the black hole \citep[][and references
  therein]{Doel_etal:12}.  This is additionally supported by
the success of semi-analytical jet models in simultaneously
reproducing the EHT observations, large-scale jet properties, and
spectral energy density (SED) of M87, that will be reported
elsewhere.  Thus, the sub-mm jet is almost certainly launched in the
immediate environment of the black hole.

All current potential mechanisms for launching relativistic jets
invoke an accretion flow.  This is trivially true for disk-launched
outflows \citep[e.g.,][]{BP}, but it is also the case for black
hole spin-powered jets \citep[e.g.,][]{BZ}.  In the latter case,
the accretion disk is needed to confine the horizon-penetrating
magnetic flux which enables the black hole's rotational energy to be
tapped.  While jet efficiencies, defined as
$\etaj\equiv\Lj/\dot{M}c^2$ where $\Lj$ is the total jet luminosity
(radiative, magnetic and mechanical), can instantaneously exceed unity
as a result of the electromagnetic tapping of the black hole spin
\citep{TNM11}, instabilities at the jet-disk interface
limit how high $\etaj$ can be in practice.  Thus, even for black hole
spin-powered jets, $\Lj$ may be used to estimate $\dot{M}$, and thus
address the existence of an event horizon in M87*.

In Section \ref{sec:LMdot} we describe how the jet power and $\dot{M}$
are related, and obtain an estimate for the latter in M87*.  
The size constraints placed by mm-VLBI are summarized in Section
\ref{sec:Size} and the relevant observational flux limits are
presented in Section \ref{sec:NIR}. The associated constraints upon
the existence of a photosphere are discussed in Section
\ref{sec:photo}. Our conclusions are summarized in Section
\ref{sec:C}.

\section{Estimates of $\dot{M}$} \label{sec:LMdot}

\subsection{Relating $\dot{M}$ to Jet Power}
All current models for launching relativistic jets require an
accretion disk. In the simplest models, the jet is merely the
innermost relativistic part of a magnetocentrifugal wind flowing out
from the surface of the disk. In such models \citep[e.g.,][]{BP}, the
jet and the lower-velocity wind are ultimately powered by 
the gravitational potential energy released by the accreting gas.
Thus the jet luminosity is limited by the overall energy efficiency of
the disk,
\begin{equation}
\eta_{\rm jet} \equiv \frac{L_{\rm jet}}{\dot{M}c^2} <\, \frac{L_{\rm
    jet}+L_{\rm wind}+L_{\rm radiation}}{\dot{M}c^2} = \eta_{\rm  disk},
\label{eq:ddjet}
\end{equation}
where the disk efficiency $\eta_{\rm disk}$ is determined by the
radius of the inner edge of the disk, i.e., the radius $r_{\rm ISCO}$
of the innermost stable circular orbit (ISCO) of the space-time
\citep{NT73},
\begin{equation}
\eta_{\rm disk} = 1 -\, \sqrt{1 - \frac{2}{3 r_{\rm ISCO}}}.
\end{equation}
The efficiency $\eta_{\rm disk}$ varies from 0.057 for a non-spinning
black hole up to 0.42 for a maximally spinning black hole. A typical
value is probably $\sim 0.1 - 0.2$. This means that, for disk-powered
jets, the mass accretion rate implied by a given jet luminosity is
roughly
\begin{equation}
\dot{M} \approx 10\, \frac{L_{\rm jet}}{c^2}. \label{etadisk}
\end{equation}

Alternatively, the jet may be powered by black hole rotation
\citep{BZ}.  In this case, the jet is launched by
large-scale ordered magnetic fields that penetrate the horizon, and
the total jet power is
\begin{equation}
\Lj = \frac{k}{4\pi c} \Omega_H^2 \Phi^2\,, \label{LBZ}
\end{equation}
where $\Omega_H\equiv a_*c/2r_+$ is the angular velocity of the
horizon, located at $r_+ = (GM/c^2)\, (1+\sqrt{1-a_*^2})$ where $a_* =
a/M$ is the dimensionless spin of the black hole, $\Phi$ is the
magnetic flux threading the horizon, and $k\approx0.045$ is a
dimensionless coefficient that varies modestly with black hole spin
(see \citealt{TNM10} for a numerical calculation of this dependence).
As Eq.~(\ref{LBZ}) shows, the jet power increases quadratically with
both black hole spin and magnetic flux. The former is limited by the
condition $a_* < 1$, and the latter is limited by the requirement that
the magnetic field through the horizon has to be confined by the
accretion flow. As a rough estimate, one could say that the accreting
gas is virialized and exerts a ram pressure of order
\begin{equation}
\Pa \approx \rho v_k^2\,,
\end{equation}
where $v_k$ is the local Keplerian velocity.  At the disk-funnel
interface, assumed to occur at a cylindrical radius $\req$, the ram
pressure must balance the magnetic pressure within the jet, and thus
\begin{equation}
\Lj 
\approx 
\frac{k\Omega_H^2}{4\pi c} \left( 2 \pi \req^2 B_{\rm eq} \right)^2 
\approx 
\frac{2 k c \pi^2 }{r_+^2}\req^4 \rho_{\rm eq} v_{k,\rm eq}^2\,,
\end{equation}
where $B_{\rm eq}$, $\rho_{\rm eq}$, and $v_{k,\rm eq}$ are the
magnetic field strength, gas density and Keplerian velocity at $\req$
and we have assumed $a_*\approx1$.

At the disk-funnel interface, the heavier gas is supported on top of
an otherwise buoyant magnetic field, and thus a balanced configuration
is generally unstable.  The interchange instability, a close relative
to the more commonly discussed Raleigh-Taylor instability, results in
the growth of gas fingers that interpenetrate the magnetosphere,
allowing accretion to occur on timescales comparable to the Keplerian
period at the interface \citep{Spruit_et_al_1995,2003ApJ...592.1042I,LN04}.  Therefore,
even in the presence of a strong funnel field, the typical accretion
rate is
\begin{equation}
\dot{M} \approx 4\pi \req^2 \rho_{\rm eq} v_{k,\rm eq}\,,
\end{equation}
differing from the previous estimate by only a factor of order unity.
As a result,
\begin{equation}
\dot{M} 
\approx
10\, \frac{L_{\rm jet}}{c^2}, 
\label{etasimple}
\end{equation}
i.e., this approximate calculation gives an estimate for $\dot{M}$ not
very different from the case of a disk-powered jet
(Eq.~\ref{etadisk}).

\citet{GA97} and \citet{LOP99} studied in greater detail the balance
between an accretion disk and a magnetic flux bundle confined around a
central black hole and concluded that, for the case of a standard
radiatively efficient thin accretion disk, the likely jet efficiency
is significantly smaller than 10\%, which means that $\dot{M}$ is
likely to be substantially larger than $10L_{\rm jet}/c^2$. The
situation is, however, different in the case of systems like M87 which
have hot advection-dominated accretion flows \citep[for a review,
  see][]{YN14}.

Recent GRMHD simulations \citep[e.g.,][]{TNM11,MTB12,NSPK12,SNPZ13}
have shown that substantial magnetic flux can be confined around a
black hole by a hot accretion flow.  The field strength is larger than
simple estimates suggest partly because of geometrical factors related
to the shape and dynamics of the accretion flow and partly because of
relativistic corrections in the vicinity of the black hole horizon
\citep{2015ASSL..414...45T}. The net result
is that the jet efficiency can be up to a factor of 10 larger than
naive estimates suggest. Indeed efficiencies exceeding 100\% are
possible if the black hole spins at close to the maximal rate and the
magnetic flux achieves its maximum value via a magnetically arrested
disk \citep[MAD,][]{NIA03,TNM11} configuration. Allowing for these
effects, we thus have
\begin{equation}
\dot{M} \gtrsim \frac{1}{2}\,\frac{L_{\rm jet}}{c^2}, \label{etaMAD}
\end{equation}
where the value we give for the coefficient is highly conservative. In
the following, we make the conservative assumption that $\dot{M} =
L_{\rm jet}/2 c^2$.

\subsection{Estimates of Jet Power in M87}

A variety of estimates of M87's jet power may be found in the
literature, covering a wide variety of distances from the black hole
and thus timescales probed. All the estimates are consistent with a
jet power of roughly $10^{44}~\ergs$.

Surrounding M87 is an extended radio-bright structure
\citep{BSS49,M52,BM54}, reaching radial distances of nearly 30~kpc,
and believed to be powered by the central AGN \citep{OEK00}.
Estimates of the jet power may be obtained by estimating the energy
required to grow the radio halo and dividing it by the buoyancy
timescale; this gives $\Lj\approx{\rm few}\,\times10^{44}~\ergs$
\citep{OEK00}.  Recent efforts to estimate the synchrotron age of the
halo using LOFAR observations give roughly 40~Myr, resulting again in
$\Lj\approx6$--$10\times10^{44}~\ergs$, depending on the assumed
particle content \citep{deGasperin_etal:12}.  These necessarily
represent estimates averaged over the buoyancy time at 10~kpc, roughly
10~Myr.  As a result, inferring the current instantaneous jet power
requires some assumption regarding the recent history of activity in
M87.

Like many radio-loud AGN, on kpc scales M87 exhibits X-ray cavities
and regions of enhanced emission associated with shocks.  These are
presumed to be driven by the AGN jet, and thus provide an additional
measure of jet power.  Estimates of the shock energetics alone require
a power source of $2.4\times10^{43}~\ergs$
\citep{Forman_etal:05,Forman_etal:07}.  The power inferred from the
nearby cavity inflation is sensitively dependent on the gas pressure
profile, and estimates range from $10^{43}$--$10^{44}~\ergs$
\citep{YWM02,Rafferty_etal:06,Allen_etal:06}. 
Again, these are time-averaged over the buoyancy time, in this case
roughly $\sim1$\,Myr because of the smaller scale (1\,kpc), with the
attendant caveats regarding variability.

Knot A is a bright optical feature within the jet at approximately
0.9~kpc.  While comparable to the distance of the closest cavities,
knot A is one of a number of superluminal features within the jet,
with apparent velocities of up to $1.6c$, implying that it moves
relativistically \citep{Meyer_etal:13}.  Interpreting knot A as an
oblique shock within the jet results in a jet power estimate of
$(1-3)\times10^{44}~\ergs$ \citep{BB96}; values much in excess
of this are expected to over-produce the emission at knot A
\citep{RFCR96}.  Equally importantly, the time delays
associated with estimates from knot A are roughly $2\times10^3~\yr$,
three orders of magnitude shorter than those associated with the
large-scale radio and X-ray morphology.

Finally, located 60~pc from M87* is the HST-1 complex, comprised of
stationary and superluminal components with apparent velocities of up
to $6c$ \citep{BSM99,Giroletti_etal:12}.  At these distances, HST-1
provides the most contemporaneous estimates of the jet power, with a
time delay of roughly 30~\yr.  If identified with a recollimation
shock, the stationary component of HST-1 implies a jet power of
$\approx10^{44}~\ergs$ \citep{Staw_etal:06}.  This is grossly
consistent with estimates obtained when the shock structure is modeled
in more detail \citep{Brom-Levi:09}.

Overall, the typical estimate for the bolometric power of M87's jet is
$\sim 10^{44}~\erg\,\s^{-1}$, with perhaps a factor of few
uncertainty.  The implied mass accretion rate is $\dot{M} \sim
10^{-3}~M_\odot\,\yr^{-1}$

\section{1.3mm VLBI Size of M87} \label{sec:Size}

The detection of emission at the core of M87 on the scale of several
times the Schwarzschild radius is robust.  Observations on an array
including telescopes in California (CARMA), Hawaii (JCMT) and Arizona
(SMT) provided projected baseline lengths ranging from
$6\times10^8\lambda$ to $3.4\times10^9\lambda$.  Firm detection of M87
on the longest baselines alone requires the existence of compact
structure with an upper size limit of $\sim60$
microarcseconds, corresponding to a diameter of $\sim16GM/c^2$ at
the presumed mass and distance of M87.  When modeled as a single
circular Gaussian brightness distribution \citep{Doel_etal:12}, the
Full Width Half Maximum is $11\pm1GM/c^2$ with calibration errors
included in the 3-$\sigma$ parameter estimate.  

Other instrumental factors can affect the estimated 1.3 mm VLBI size
but were determined to be negligible.  These include the stability of
Hydrogen Maser frequency standards used at each VLBI station and the
polarization purity of the telescope receivers which were configured
for Left Circular Polarization.  On short ($\approx1$~s) timescales, the
Masers were compared to ultra-stable quartz oscillators, and
longer-term stability was determined through timing comparisons with
GPS.  Polarization characteristics at each antenna were determined by
injecting known polarization signals into the receiver feeds at the
single dish sites, and through separate polarization calibration
observations at CARMA. 

The main source of uncertainty in the true size
and shape of the compact 1.3mm emission is the limited VLBI baseline
coverage, which is insufficient to exactly determine the brightness
morphology.  The VLBI data can also be fitted, for example, with a
uniform disk of diameter $16.4GM/c^2$, comparable to that
inferred from the the long-baseline correlated flux measurements
alone.  In this work we adopt the circular Gaussian
diameter of $11GM/c^2$ with the proviso that future EHT observations
with improved Global coverage will be able to image and model M87 in
detail.  The disk and more complex models typically predict ``nulls'' in
the VLBI signal as a function of baseline length that are not yet
seen, but which may emerge as EHT sites enabling longer baselines are
added to the array.

\citet{Doel_etal:12} associate the 1.3mm compact emission with the
ISCO of the M87 black hole, enlarged by the strong gravitational
lensing that occurs at small radii.  Location of this emission close to, or
at, the black hole is supported by phase referenced multi-frequency
VLBI observations at longer wavelengths, which show the core of M87 to
shift towards a convergent point with shorter observing wavelength
\citep{2011Natur.477..185H}.  Extrapolation of this shift to the observed size
at 1.3mm places the emission within a few Schwarzschild radii
of the black hole, making the derived 1.3mm size a good estimate
for the photosphere of a putative surface.  

While it is possible that the 1.3mm emission arises some distance from
the black hole, the EHT-derived size can still serve as an upper limit
on the spatial extent of an intrinsic photosphere.  
Theoretical models of electromagnetic jets, both semi-analytical
estimates and numerical simulations, uniformly show an expanding jet
width profile with increased distance from the central
engine.  These expectations are quantitatively consistent with jet
observations on scales covering six orders of magnitude
\citep{2013ApJ...775...70H,2013ApJ...775..118N} 
confirm a parabolic jet shape to within $\sim20GM/c^2$ of the black
hole.  There is, in short, no compelling mechanism to suggest that 
an electromagnetic jet launched from a photosphere later converges to
a markedly smaller size.

\section{The SED of the Nuclear AGN Source in M87} \label{sec:NIR}

\begin{deluxetable*}{cccccccc}
\tablecaption{Ultravoilet--Infrared SED of the M87 Nuclear Source \label{tab:Fnu}}
\tablehead{
$\lambda$ & $\log{\nu}$  & $F_\nu$ & $\sigma_F$ & Epoch & Program & Camera & Source\mbox{\hspace{1.5in}}\\
($\mu$m)  & (Hz)        & (mJy)   & (mJy)      &       & HST ID      &               &                \\
(1)       & (2)         & (3)     & (4)        & (5)   & (6)         & (7)           & (8)            }
\startdata
     0.1255    & 15.38    & 0.162   & 0.032      & 1991 Apr  6 & 1228 & FOC           & \citet{Spar-Bire-Macc:96}                \\
     0.1507    & 15.30    & 0.081   & 0.016      & 1991 Apr  5 & 1228 & FOC           & \citet{Spar-Bire-Macc:96}                \\
     0.1585    & 15.28    & 0.068   & 0.014      & 1991 Jun 23 & 1517 & FOC           & \citet{Spar-Bire-Macc:96}                \\
     0.23      & 15.11    & 0.158   & 0.032      & 1991 Apr  5 & 1228 & FOC           & \citet{Spar-Bire-Macc:96}                \\
     0.25      &          & 0.1153  & 0.0014     & 2003 Mar 31 & 9454 & ACS HRC       & \citet{Maoz_etal:05}                  \\
     0.25      &          & 0.1119  & 0.0014     & 2003 Dec 10 & 9454 & ACS HRC       & \citet{Maoz_etal:05}                  \\
     0.2475    &          & 0.089   & 0.009      & 1999 May 17 & 8140 & STIS NUV-MAMA & \citet{Chia_etal:02}             \\
     0.299     &          & 0.198   & 0.005      & 1998 Feb 25 & 6775 & WFPC2 PC      & This paper; \citet{Perl_etal:01a}   \\ % F300W
     0.33      &          & 0.1759  & 0.0019     & 2003 Mar 31 & 9454 & ACS HRC       & \citet{Maoz_etal:05}                  \\
     0.33      &          & 0.1743  & 0.0019     & 2003 Dec 10 & 9454 & ACS HRC       & \citet{Maoz_etal:05}                  \\
     0.3708    & 14.91    & 0.457   & 0.091      & 1991 Jun 23 & 1517 & FOC           & \citet{Spar-Bire-Macc:96}                \\
     0.456     &          & 0.493   & 0.012      & 1998 Feb 25 & 6775 & WFPC2 PC      & This paper; \citet{Perl_etal:01a}   \\ % F450W
     0.5017    & 14.78    & 1.02    & 0.20       & 1991 Apr  5 & 1228 & FOC           & \citet{Spar-Bire-Macc:96}                \\
     0.541     &          & 0.637   & 0.059      & 1991 Feb 24 & 1105 & WF/PC PC      & \citet{Laue_etal:92}; this paper     \\
     0.600     &          & 0.787   & 0.020      & 1998 Feb 25 & 6775 & WFPC2 PC      & This paper; \citet{Perl_etal:01a}   \\ % F606W
     0.801     &          & 1.063   & 0.027      & 1998 Feb 25 & 6775 & WFPC2 PC      & This paper; \citet{Perl_etal:01a}   \\ % F814W
     0.801     &          & 0.759   & 0.076      & 1999 May 11 & 8140 & WFPC2         & \citet{Chia-Cape-Celo:99,Chia_etal:02}    \\ % F814W
     0.8900    &          & 0.81    & \dots      & 1991 Jun  1 & 3242 & WF/PC PC      & \citet{Laue_etal:92}                 \\
     1.2       &          & 2.45    & 0.69       & 1993 Jun  1 & \dots& ESO/MPI 2 m   & \citet{Spar-Bire-Macc:96}             \\
     1.60      &          & 3.02    & 0.15       & 1997 Nov 20 & 7171 & NICMOS NIC2   & \citet{Baldi_etal:10}                 \\ % Rieke
     2.201     &          & 2.28    & 0.60       & 1994 Apr  3 & \dots& UKIRT 3.8 m   & \citet{Spar-Bire-Macc:96}             \\
\enddata
\tablecomments{NOTES -- Columns (1) and (2) are the effective, pivot, or median wavelength and frequency of the photometric 
         bandpass.  Columns (3) and (4) are the measured flux and its 1-$\sigma$ error.
         Because the nuclear source flux is time-variable, Column (5) gives the epoch of
         the observation.  This is not used in the present paper: all SED points are 
         plotted in Figure~1 to provide an illustration of the amplitude of variation.
         This amplitude is in general larger than the errors of the SED measurements.
         Column (6) provides the HST proposal ID, and
         Column (7) lists the HST camera or the telescope used.
         Column (8) gives the source of the measurement. 
         }
\end{deluxetable*}

\begin{deluxetable*}{cccclrc}
\tablecaption{Ultravoilet--Optical SED of the M87 Nuclear Source from \citet{Perl_etal:11} \label{tab:Fnu2}}
\tablehead{
 $F_\nu$ (mJy)     & $F_\nu$ (mJy)      & $F_\nu$ (mJy)     & $F_\nu$ (mJy)        & Epoch       & Program & Camera   \\
 F606W             & F330W              & F250W             & F220W                &             & HST ID  &          \\
 0.589 $\mu$m      & 0.336 $\mu$m       & 0.272 $\mu$m      & 0.226 $\mu$m         &             &         &          \\
 (1)               & (2)                & (3)               & (4)                  & (5)         & (6)     & (7)      }
\startdata
$ 0.671 \pm 0.007 $&$  0.305 \pm 0.002 $&$        \dots     $&$   0.146 \pm 0.011 $& 2002 Dec  7 & 9705   & ACS HRC  \\
$ 0.630 \pm 0.006 $&$  0.305 \pm 0.002 $&$        \dots     $&$   0.146 \pm 0.011 $& 2002 Dec 10 & 9493   & ACS HRC  \\
$ 0.478 \pm 0.005 $&$        \dots     $&$        \dots     $&$   0.137 \pm 0.011 $& 2003 Nov 29 & 9829   & ACS HRC  \\
$ 1.066 \pm 0.011 $&$  0.475 \pm 0.002 $&$  0.306 \pm 0.014 $&$   0.226 \pm 0.014 $& 2004 Nov 28 & 10133   & ACS HRC  \\
$ 1.306 \pm 0.013 $&$        \dots     $&$  0.363 \pm 0.010 $&$         \dots     $& 2004 Dec 26 & 10133   & ACS HRC  \\
$ 0.891 \pm 0.009 $&$        \dots     $&$  0.280 \pm 0.009 $&$         \dots     $& 2005 Feb  9 & 10133   & ACS HRC  \\
$ 1.037 \pm 0.010 $&$        \dots     $&$  0.328 \pm 0.010 $&$         \dots     $& 2005 Mar 27 & 10133   & ACS HRC  \\
$ 0.932 \pm 0.009 $&$  0.446 \pm 0.003 $&$  0.274 \pm 0.009 $&$   0.217 \pm 0.009 $& 2005 May  9 & 10133   & ACS HRC  \\
$ 0.839 \pm 0.008 $&$        \dots     $&$  0.273 \pm 0.009 $&$         \dots     $& 2005 Jun 22 & 10133   & ACS HRC  \\
$ 0.639 \pm 0.006 $&$        \dots     $&$  0.192 \pm 0.007 $&$         \dots     $& 2005 Aug  1 & 10133   & ACS HRC  \\
$ 0.735 \pm 0.007 $&$  0.349 \pm 0.002 $&$  0.217 \pm 0.008 $&$   0.170 \pm 0.008 $& 2005 Nov 29 & 10133   & ACS HRC  \\
$ 0.756 \pm 0.008 $&$        \dots     $&$  0.234 \pm 0.008 $&$         \dots     $& 2005 Dec 26 & 10617   & ACS HRC  \\
$ 0.631 \pm 0.006 $&$        \dots     $&$  0.201 \pm 0.008 $&$   0.160 \pm 0.008 $& 2006 Feb  8 & 10617   & ACS HRC  \\
$ 0.780 \pm 0.008 $&$        \dots     $&$  0.232 \pm 0.008 $&$         \dots     $& 2006 Mar 30 & 10617   & ACS HRC  \\
$ 0.862 \pm 0.009 $&$  0.372 \pm 0.002 $&$  0.193 \pm 0.007 $&$  0.1564 \pm 0.007 $& 2006 May 23 & 10617   & ACS HRC  \\
$ 1.370 \pm 0.014 $&$  0.636 \pm 0.003 $&$  0.323 \pm 0.009 $&$         \dots     $& 2006 Nov 28 & 10910   & ACS HRC  \\
$ 1.006 \pm 0.010 $&$        \dots     $&$  0.276 \pm 0.009 $&$         \dots     $& 2006 Dec 30 & 10910   & ACS HRC  \\
$ 1.292 \pm 0.017 $&$        \dots     $&$        \dots     $&$         \dots     $& 2007 Nov 25 & 11216   & WFPC2 WFC \\
\enddata
\tablecomments{Notes -- The first four columns list fluxes $F_\nu$
  (mJy) for the four filters and their pivot wavelengths as listed in
  the table headers.  Columns (5) -- (7) are explained in their
  headers.}
\end{deluxetable*}

Tables \ref{tab:Fnu} and \ref{tab:Fnu2} list the optical and
near-infrared SED data that are used in Figures \ref{fig:spec} and
\ref{fig:fr}.  Table \ref{tab:Fnu2} lists the measurements~made~in the
\citet{Perl_etal:11} five-year campaign to measure nuclear
variability.  Table \ref{tab:Fnu} lists the rest of the data.

Most flux measurements were made with the {\it Hubble Space Telescope}
({\it HST\/}) and various imaging cameras.  Also, unless otherwise
noted, nuclear flux measurements were made using a correction for the
galaxy light under the central PSF that is based on measuring and
fitting the profile at larger radii 
\citep[see, e.g.,][]{Laue_etal:92,Spar-Bire-Macc:96}.   Various
profile fit models such as S\'ersic functions \citep{Sers:68} and ones
with a core \citep[see][]{Laue_etal:95} are used.  Also unless
otherwise noted, the innermost jet knot HST-1 is not included in the
measurements, although faint extensions of the jet at smaller radii
are included \citep[see, e.g.,][]{Spar-Bire-Macc:96,Perl_etal:01a}.
The measurements are discussed in their source papers.  A few details
are noteworthy here: 

An estimate by \citet{Youn_etal:78} of $V \simeq 16.69 \pm 0.05$
\citep[see][]{Laue_etal:92} corresponding to $F_\nu = 0.80 \pm
0.04$~mJy is consistent with the results in Table \ref{tab:Fnu} but is
not used in this paper.  The epoch was 1975~Mar~14 -- 1977~May~22 and
the Palomar Observatory 200-inch Hale telescope was used.  We omit
this measurement because the seeing FWHM was 1$^{\prime\prime}$ to
2$^{\prime\prime}$ -- hence the exclusion of jet knots is uncertain --
and because the galaxy subtraction was not consistent with our HST-era
understanding of the cuspy cores of elliptical galaxies
\citep[e.g.,][]{Laue_etal:92,Laue_etal:95}. 

The \citet{Laue_etal:92} HST PC F785LP images 
($\lambda_{\rm eff}=8900$~\AA) were taken with the unrefurbushed HST
but allow a clear separation of the nuclear AGN from the innermost jet
knots N1 and M at $\sim$0\sd1 and $\sim$~0\sd18 from the center (see
their Figure 5).  These knots are included in most HST photometry and
all ground-based photometry.  \citet{Laue_etal:92} measure them
separately and show that, at 8900~\AA~and on 1999 Jun 1, the nucleus
contributed 89\% of the combined flux including N1 and M.  We have
used their measurement of the nuclear flux at 8900~\AA.  They also
measure the total flux of the nucleus plus N1 plus M in the F555W band
at almost the same time as the F785LP measurements.  We assume that
the fraction of this total that comes from the nucleus is also 89\%
and correct their total measurement to a nuclear estimate of
0.637~mJy.  {\it This correction has not been applied to any other
  measurements, because we cannot know the variability of the nucleus,
  of N1, and of M separately.  All these other measurements include
  the innermost part of the jet interior to knot HST-1.} 

We repeated the F55W galaxy subtraction to get the estimated error
$\pm 0.059$ on the F555W flux.

The \citet{Spar-Bire-Macc:96} measurements were made with pre-COSTAR
{\it HST\/} using the ground-based photometry by \citet{Youn_etal:78}
as the galaxy model.  The latter is probably not a problem; they note
that corrections for galaxy light are small.  If the former is a
problem, then light at large radii in the PSF could be missed and the
measurements could be lower limits on fluxes.  However, we emphasize
that this problem is almost certainly much smaller than the intrinsic
variability amplitude of the nuclear source.

\citet{Spar-Bire-Macc:96} measured the nuclear AGN in $J$ and $K$
bands using the ESO/MPI~2~m telescope and the United Kingdom Infrared
Telescope (UKIRT: 3.8 m), respectively.  Galaxy light underlying the
AGN was removed using an iterative procedure to separate the stellar
brightness profile from the central source.  That source was measured
inside a 2\sd8-diameter aperture, i.e., one that is large enough to
include jet knot HST-1.  The knot is not visible in their image, and
the higher-resolution HST images in \citet{Spar-Bire-Macc:96} suggest
that it contributed negligibly at that time.  There are jet components
closer to the center \citep[e.g.,][]{Laue_etal:92}; these are included
even in most HST measurements.  We therefore use the
\citet{Spar-Bire-Macc:96} measurements as published. 

The \citet{Chia-Cape-Celo:99,Chia_etal:02} measurements were made by
subtracting from the nuclear flux the galaxy~flux at radius 0\sd23 in
the $R$ band and at radius 0\sd17 in the ultraviolet band.  Since the
galaxy's profile is a shallow power law inside the cuspy core
\citep{Laue_etal:92}, the resulting fluxes are strictly speaking upper
limits.  However, the galaxy contribution to the measurements is
smaller than the amplitude of the variability of the M~87 nuclear
source.

The four HST WFPC2 points from this paper were measured by
E.S.P. following the procedures of \citet{Perl_etal:01a} in February
of 1998.  Knot HST-1 was about 1\% of the central source at this time
and  is not included in the measurement.  Prior to measuring the
nuclear flux, a galaxy model (made with the IRAF tasks \verb+ellipse+
and \verb+bmodel+) was subtracted from each of these images.  The
\citet{Perl_etal:11} measurements listed in Table \ref{tab:Fnu2} were
made in the same way.

We have not included the UV photometry of the nucleus published by
\citet{Madr:09} because all of those fluxes have been referenced to a
single wavelength based on an unpublished spectral model.  Instead, we
use the measurements published by \citet{Perl_etal:11}, which include
all of the epochs between 2004-2006 that were included in
\citet{Madr:09} but does not apply any spectral model in their
calibration.  It  should be noted that when the spectral referencing
applied by \citet{Madr:09} is removed, the measurements of
\citet{Perl_etal:11} and \citet{Madr:09} agree to within the errors
\citep{Madr:PC}.

\section{Photosphere Emission} \label{sec:photo}

Were a photosphere present in M87* instead of a horizon, it would be
heated by the constant deposition of kinetic energy from the
accretion, resulting in an additional component in the spectrum.
Since the infall time is only logarithmically dependent upon the
photosphere redshift, it is natural to assume that the surface will
reach a steady state in which the emitted luminosity and the impinging
kinetic power from the accretion flow are balanced.

The gravitational binding energy liberated per unit time by the
infalling material necessarily depends on the size of the putative
photosphere, approaching $\dot{M} c^2$ for radii comparable to that
of the horizon.  That is, typically,
\begin{equation}
L_{\rm surf} \approx \frac{GM}{c^2R}\,\dot{M} c^2\,,
\end{equation}
where $R$ is the radius of the surface.\footnote{While quantitative
  variations from the Newtonian expressions do occur for high-redshift
  objects, these are qualitatively similar, differing by a factor of
  order unity that depends primarily on the dynamical state of the
  infalling gas \citep[see, e.g.,][and Equation
    \ref{etadisk}]{BLN09}.}

The radius of any such surface is strongly limited by the 1.3mm VLBI
observations discussed in Section \ref{sec:Size}.  These already
constrain the 1.3mm emission region to within the vicinity of the
photon orbit, expected to have a diameter of roughly 
$10.4 GM/c^2$ in projection and depending very weakly on the assumed
black hole spin.\footnote{While it is true that the projected photon
  orbit viewed from the equatorial plane can shrink to $9GM/c^2$ for
  maximally spinning black holes, this is not the case at small
  inclinations; at the estimated inclination of $<25^\circ$ of the
  M87*'s jet \citep{Hein-Bege:97}, and thus presumably spin
  inclination, spin has negligible impact on the projected photon
  orbit size.}  Moreover, at millimeter wavelengths and longer M87
still exhibits the flat radio spectrum indicative of a self-absorbed
synchrotron jet \citep[see, e.g.,][]{Blan-Koni:79}.  This
interpretation receives strong support from the evolution of the radio
core location with wavelength, with the offset evolving
$\propto\lambda^{0.94\pm0.09}$, presumably due to the shrinking of the
photosphere \citep{2011Natur.477..185H}.  Thus, at infrared and
optical wavelengths the size of the photosphere is expected to be
smaller than the $\sim11GM/c^2$ limit from the mm-VLBI observations
described in Section \ref{sec:Size} -- if the optically thick to thin
transition completes at a wavelength of 1.1mm or shorter, this will
lie within the projected photon orbit.  This conclusion is only
strengthened if the emission arises from a significant distance from
the central engine, as described in Section \ref{sec:Size}.

However, the efficiency with which the jet is driven depends on the
surface in a similar fashion, becoming larger for more compact
objects.  For disk driven jets, the surface and jet efficiences are
approximately equal (see Equation \ref{eq:ddjet}), implying that
$L_{\rm surf}\approx\Lj$.  For black hole rotation driven
jets the efficiency scales roughly as $v_k$ due to the competition
between the lower $\Omega_H$ and higher $\Phi$ as the surface
increases in size (see the Appendix).  As a result, once again
$L_{\rm surf}\approx\Lj$, with an additional factor of order unity
that depends only very weakly on the photosphere size.  Thus, here we
make the conservative assumption that $L_{\rm surf}\approx\Lj/2$ as
described in Equation \ref{etaMAD}.

Since the size constraints placed by EHT observations at 1.3mm 
already limit the size of any putative photospheric emission in M87*
to scales comparable to, or smaller than, the photon orbit, the
photosphere will necessarily be thermalized by the strong lensing,
resulting in a gravitational analog of the standard blackbody cavity 
\citep[see the discussion in ][]{BN06,BN07}. Thus, surface should be 
temperature as seen by a distant observer is related to the accretion
rate via
\begin{equation}
4\pi R_a^2 \sigma T_\infty^4 \approx \frac{\Lj}{2c^2}
\quad\rightarrow\quad
T_\infty \approx \left( \frac{\Lj}{8\pi R_a^2 \sigma} \right)^{1/4}\,,
\label{eq:Tinf}
\end{equation}
where $\sigma$ is the Stefan-Boltzmann constant and $R_a\approx 5
GM/c^2$ is the apparent radius of any object lying within the photon
orbit.  Note that the temperature estimate depends only weakly on
$\dot{M}$, and thus on the uncertainties associated with the jet power
and jet efficiency.  For the accretion rate inferred in Section
\ref{sec:LMdot} for M87*, $T_\infty\approx8.5\times10^3~\K$, which
means that the surface emission will peak in the optical band.  The
observed spectral flux from the surface is given by
\begin{equation}
F_{\nu,\rm ph} 
=
\frac{R_a^2}{D^2} \frac{2h\nu^3/c^2}{e^{h\nu/kT_\infty}-1}\,,
\end{equation}
where $D$ is the distance to M87.  This predicted flux may be directly
compared to observations and thereby we can hope to constrain whether
or not M87* has a surface. 

\section{Discussion} \label{sec:C}

\begin{figure}
\begin{center}
\includegraphics[width=\columnwidth]{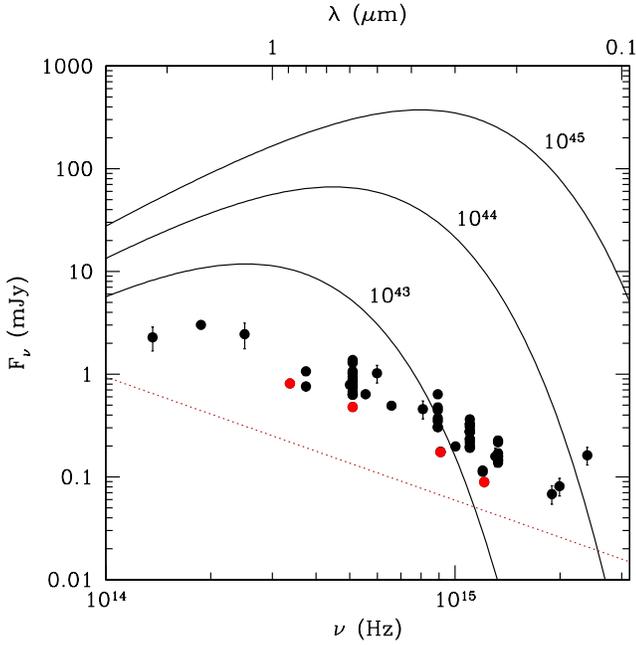}
\end{center}
\caption{Inferred infrared-optical spectrum from a putative
  photosphere for $L_{\rm photo}=10^{43}~\erg\,\s^{-1}$,
  $10^{44}~\erg\,\s^{-1}$, and $10^{45}~\erg\,\s^{-1}$ (solid lines,
  from bottom to top) compared to the empirical limits on the SED of
  M87* listed in Tables \ref{tab:Fnu} and \ref{tab:Fnu2} (colored
  dots).  The dark-red dotted line indicates the estimated limit
  arising from the intrinsic jet spectrum, and thus presumably the
  confusion limit.  All points in Table \ref{tab:Fnu2} are plotted in
  order to provide some indication of the variability of the
  source. Those data that provide the constraints presented in Figure
  \ref{fig:fr} are shown in red. Note that photospheric emission at a
  level of $10^{44}~{\rm erg\,s^{-1}}$, as expected based on current
  estimates of the jet power (\S2), is ruled out with high
  confidence.} \label{fig:spec} 
\end{figure}

\begin{figure}
\begin{center}
\includegraphics[width=\columnwidth]{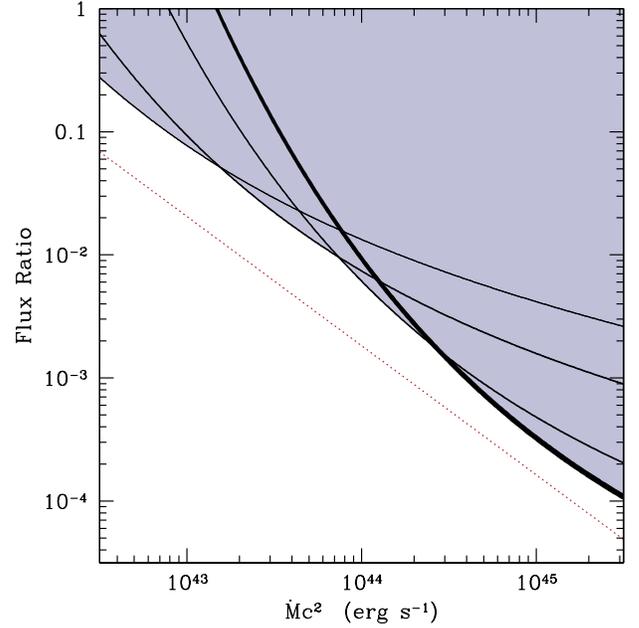}
\end{center}
\caption{Ratio of the measured fluxes to that inferred from a putative
  photosphere.  The thick black lines indicate the constraint implied
  by the individual flux measurements shown in red in Fig
  \ref{tab:Fnu}, corresponding to those that dominate the constraint
  at some $\dot{M}$ in the range presented.  The line widths indicate
  the 1$\sigma$ upper limit and the grayed region is the area excluded
  by the aggregate.  The dark-red dotted line indicates the estimated
  limit arising from the intrinsic jet spectrum, and thus presumably
  the confusion limit.}\label{fig:fr}
\end{figure}

A direct comparison of the putative photosphere spectrum to the
measured fluxes listed in Table \ref{tab:Fnu} is shown in Figure
\ref{fig:spec}.  Because of the intrinsic uncertainty in $\dot{M}$, we
show predicted spectra spanning a wide range of $\dot{M}$,
corresponding to variations in the jet efficiency or measured jet
power.  In all cases, photospheric emission is excluded at high
confidence.  This conclusion is further supported by Figure
\ref{fig:fr} which shows the ratio of the measured fluxes to that
implied by a photosphere as a function of $\dot{M}$.  At all
physically reasonable accretion rates the measured fluxes lie at least
an order of magnitude below that required in the absence of an event
horizon.

Improvements in the measured flux limits will necessarily result in
increased confidence with which any photospheric emission may be ruled
out.  However, the intrinsic emission from the observed jet component
provides a natural limit.  Assuming a spectral index of 1.2,
consistent with a transition from optically thick to optically thin
near 1mm, and the optical flux limits, we estimate the 
corresponding floors on the infrared-optical spectrum and flux ratio,
shown by the dotted, dark-red lines in figures \ref{fig:spec} and
\ref{fig:fr}, below which the emission from the jet-launching region
must be modeled in detail.  Note that this is rather uncertain; variations in
the spectral index between the core and larger scale jet emission
arising, e.g., from an evolution in the relativistic particle
distributions, could easily harden the core spectrum sufficiently to
fully account for the vast majority of the fluxes shown in Figure
\ref{fig:spec}.  Nevertheless, even if this is not the case, it
appears that the current flux limits are within an order of magnitude
of the confusion limit, below which careful modeling of the jet
emission will be required to obtain stronger constraints.  

Despite the astrophysical uncertainty in the relationship between the
jet power and mass accretion rate, we have shown in this paper that,
within the context of current jet launching paradigms, we can rule out
the existence of an observable photosphere in M87* within which the
kinetic energy of the accreting gas is deposited.  The implication is
that the kinetic energy of the gas is advected past an event horizon,
beyond which it is no longer visible to distant observers. In other
words, M87* must have an event horizon.

This argument may be evaded in a variety of ways, though all require
speculative new physics and most invoke black hole
alternatives. First, M87* could fail to reach steady state, preventing
the use of the accretion rate as a proxy for the surface luminosity.
This requires unphysically high heat capacities (which the gravastar
model of \citealt{Mazur_Mottola_2004} apparently does possess, see
\citealt{Chapline_2005}), already excluded in many cases \citep{BN07},
or exotic alternatives like suitably tuned wormholes
\citep[e.g.,][]{DS07}.  Second, the surface could fail to thermalize,
though this is strongly argued against by the compact emission
observed at millimeter wavelengths, and may be excluded altogether by
the development of a baryonic atmosphere.  Third, the efficiency
factor could be much larger than implied by current jet launching
models, requiring mechanisms qualitatively different from those
currently favored.  For example, if M87* were a rapidly spinning
compact object with a strong magnetic field, the accreting gas could
be stopped at a magnetospheric radius and flung out in a jet without
any gas reaching the underlying photosphere (e.g., the neutron star
propeller model of \citealt{Illarionov_Sunyaev_1975}). In any such
model, the magnetospheric radius can be only slightly larger than the
surface of the central object (because of the VLBI constraints on the
angular size of the source), and it seems likely that a lot of
accreting gas will penetrate through the magnetosphere and reach the
surface (e.g., \citealt{Kulkarni_Romanova_2008}). In addition, the
material that is stopped and flung out by the magnetosphere is likely
to dissipate a good fraction of its kinetic energy where it meets the
rotating magnetosphere, resulting in thermalized radiation not very
different from that expected from a photosphere.  We regard this
propeller model as strictly physically possible but astrophysically
implausible.  Thus, our argument will be strengthened in the near
future as key components to the current jet paradigm are critically
tested by mm-VLBI observations of M87.

\acknowledgments 
Based on observations made with the Event Horizon Telescope and the
NASA/ESA {\it Hubble Space Telescope}.  The Event Horizon Telescope is
supported through grants from the US National Science Foundation, by
the Gordon and Betty Moore Foundation (GBMF3561), and through
generous equipment donations from the Xilinx Corporation. 
The NASA/ESA {\it Hubble Space Telescope} observations were obtained at the
Space Telescope Science Institute, which is operated by AURA, Inc., under NASA contract NAS
5-26555.  These observations are associated with program numbers 1105,
1228, 1517, 3242, 6775, 7171, 8140, 9454, 9493, 9705, 9829, 10133,
10617, 10910, and 11216.  A.E.B.~receives financial support from
Perimeter Institute for Theoretical Physics and the Natural Sciences
and Engineering Research Council of Canada through a Discovery
Grant. RN received partial supported from the NSF via grant AST1312651
and NASA via grant NNX14AB47G. RN also thanks the Perimeter Institute
for hospitality while some of this work was carried out.  The authors
are grateful to Sera Markoff and Brian McNamara for helpful
conversations.  This work would not have been practical without
extensive use of the NASA/IPAC Extragalactic Database (NED), which is
operated by the Jet Propulsion Laboratory and the California Institute
of Technology under contract with NASA.  We also made extensive use of
NASA's Astrophysics Data System bibliographic services.  JK's work was
supported in Texas by the Curtis T.~Vaughan, Jr.~Centennial Chair in
Astronomy and in Germany by a Faculty Research Assignment from the
University of Texas and by the Max-Planck-Institut f\"ur
Extraterrestrsiche Physik (MPE), Garching-by-Munich, Germany.  JK
warmly thanks Director Ralf Bender and the staffs of the MPE and the
Universit\"ats-Sternwarte, Ludwig-Maximilians-Universit\"at, Munich
for their hospitality and support during the 2014 visit when most of
his work on this paper was done.

Facilities:
\facility{EHT: James Clerk Maxwell Telescope, Arizona Radio Observatory Submillimeter Telescope, Combined Array for Research in mm-wave Astronomy, the Submillimeter Array}
\facility{HST: FOC},
\facility{HST: WFPC1},
\facility{HST: WFPC2},
\facility{HST: NICMOS},
\facility{HST: ACS},
\facility{HST: STIS},
\facility{UKIRT 3.8 m telescope},
\facility{ESO/MPI 2 m telescope},

\begin{appendix}
\section{Blandford-Znajek Jet Efficiency for Surfaces}
The relationships between accretion rate and both surface luminosity
and jet power are dependent upon the assumed surface size.  In detail
this depends upon the specific jet launching mechanism assumed.  Here
we consider the question in the context of Blandford-Znajek jets, for which
the jet luminosity is related to the properties of the central object
via Equation \ref{LBZ}.  For an object with a large surface there
are two important modifications: the angular velocity is limited to
$\Omega_H<c/R$, and $\Phi<2\pi R^2 B_{\rm eq}$.  Thus,
\begin{equation}
\Lj 
\lesssim 
\pi k c R^2 B_{\rm eq}^2
\approx
2\pi k \frac{v_{k,\rm eq}}{c} \dot{M} c^2\,,
\end{equation}
where we assumed $\dot{M}\approx4\pi R^2\rho_{\rm eq} v_{k,\rm eq}$
and equipartion near the surface (i.e., $\req\approx R$).  This must
be compared to
\begin{equation}
L_{\rm surf} \approx \frac{GM}{c^2 R} \dot{M} c^2 \approx \dot{M} v^2_{k,\rm eq}\,,
\end{equation}
which gives
\begin{equation}
\frac{L_{\rm surf}}{\Lj}
\approx
\frac{1}{2\pi k}
\frac{v_{k,\rm eq}}{c}
\propto
\left(\frac{GM}{c^2 R}\right)^{1/2}\,.
\end{equation}
This is a weak function of $R$. For the range of surface radii
consistent with the recent mm-VLBI limits, it introduces at most a factor
of two.
\end{appendix}

\bibliography{jss.bib,m87.bib}

\begin{thebibliography}{}
\expandafter\ifx\csname natexlab\endcsname\relax\def\natexlab#1{#1}\fi

\bibitem[{{Allen} {et~al.}(2006){Allen}, {Dunn}, {Fabian}, {Taylor}, \&
  {Reynolds}}]{Allen_etal:06}
{Allen}, S.~W., {Dunn}, R.~J.~H., {Fabian}, A.~C., {Taylor}, G.~B., \&
  {Reynolds}, C.~S. 2006, \mnras, 372, 21

\bibitem[{{Almheiri} {et~al.}(2013){Almheiri}, {Marolf}, {Polchinski}, \&
  {Sully}}]{2013JHEP...02..062A}
{Almheiri}, A., {Marolf}, D., {Polchinski}, J., \& {Sully}, J. 2013, Journal of
  High Energy Physics, 2, 62

\bibitem[{{Baade} \& {Minkowski}(1954)}]{BM54}
{Baade}, W., \& {Minkowski}, R. 1954, \apj, 119, 215

\bibitem[{{Baldi} {et~al.}(2010){Baldi}, {Chiaberge}, {Capetti}, {Sparks},
  {Macchetto}, {O'Dea}, {Axon}, {Baum}, \& {Quillen}}]{Baldi_etal:10}
{Baldi}, R.~D., {Chiaberge}, M., {Capetti}, A., {et~al.} 2010, \apj, 725, 2426

\bibitem[{{Batcheldor} {et~al.}(2010){Batcheldor}, {Robinson}, {Axon},
  {Perlman}, \& {Merritt}}]{Batc_etal:10}
{Batcheldor}, D., {Robinson}, A., {Axon}, D.~J., {Perlman}, E.~S., \&
  {Merritt}, D. 2010, \apjl, 717, L6

\bibitem[{{Bicknell} \& {Begelman}(1996)}]{BB96}
{Bicknell}, G.~V., \& {Begelman}, M.~C. 1996, \apj, 467, 597

\bibitem[{{Biretta} {et~al.}(1999){Biretta}, {Sparks}, \& {Macchetto}}]{BSM99}
{Biretta}, J.~A., {Sparks}, W.~B., \& {Macchetto}, F. 1999, \apj, 520, 621

\bibitem[{{Blandford} \& {K{\"o}nigl}(1979)}]{Blan-Koni:79}
{Blandford}, R.~D., \& {K{\"o}nigl}, A. 1979, \apj, 232, 34

\bibitem[{{Blandford} \& {Payne}(1982)}]{BP}
{Blandford}, R.~D., \& {Payne}, D.~G. 1982, \mnras, 199, 883

\bibitem[{{Blandford} \& {Znajek}(1977)}]{BZ}
{Blandford}, R.~D., \& {Znajek}, R.~L. 1977, \mnras, 179, 433

\bibitem[{{Bolton} {et~al.}(1949){Bolton}, {Stanley}, \& {Slee}}]{BSS49}
{Bolton}, J.~G., {Stanley}, G.~J., \& {Slee}, O.~B. 1949, \nat, 164, 101

\bibitem[{{Broderick} {et~al.}(2009){Broderick}, {Loeb}, \& {Narayan}}]{BLN09}
{Broderick}, A.~E., {Loeb}, A., \& {Narayan}, R. 2009, \apj, 701, 1357

\bibitem[{{Broderick} \& {Narayan}(2006)}]{BN06}
{Broderick}, A.~E., \& {Narayan}, R. 2006, \apjl, 638, L21

\bibitem[{{Broderick} \& {Narayan}(2007)}]{BN07}
---. 2007, Classical and Quantum Gravity, 24, 659

\bibitem[{{Bromberg} \& {Levinson}(2009)}]{Brom-Levi:09}
{Bromberg}, O., \& {Levinson}, A. 2009, \apj, 699, 1274

\bibitem[{{Chapline}(2005)}]{Chapline_2005}
{Chapline}, G. 2005, in 22nd Texas Symposium on Relativistic Astrophysics, ed.
  P.~{Chen}, E.~{Bloom}, G.~{Madejski}, \& V.~{Patrosian}, 101--104

\bibitem[{{Chiaberge} {et~al.}(1999){Chiaberge}, {Capetti}, \&
  {Celotti}}]{Chia-Cape-Celo:99}
{Chiaberge}, M., {Capetti}, A., \& {Celotti}, A. 1999, \aap, 349, 77

\bibitem[{{Chiaberge} {et~al.}(2002){Chiaberge}, {Macchetto}, {Sparks},
  {Capetti}, {Allen}, \& {Martel}}]{Chia_etal:02}
{Chiaberge}, M., {Macchetto}, F.~D., {Sparks}, W.~B., {et~al.} 2002, \apj, 571,
  247

\bibitem[{{Damour} \& {Solodukhin}(2007)}]{DS07}
{Damour}, T., \& {Solodukhin}, S.~N. 2007, \prd, 76, 024016

\bibitem[{{de Gasperin} {et~al.}(2012){de Gasperin}, {Orr{\'u}}, {Murgia},
  {Merloni}, {Falcke}, {Beck}, {Beswick}, {B{\^i}rzan}, {Bonafede},
  {Br{\"u}ggen}, {Brunetti}, {Chy{\.z}y}, {Conway}, {Croston}, {En{\ss}lin},
  {Ferrari}, {Heald}, {Heidenreich}, {Jackson}, {Macario}, {McKean}, {Miley},
  {Morganti}, {Offringa}, {Pizzo}, {Rafferty}, {R{\"o}ttgering}, {Shulevski},
  {Steinmetz}, {Tasse}, {van der Tol}, {van Driel}, {van Weeren}, {van
  Zwieten}, {Alexov}, {Anderson}, {Asgekar}, {Avruch}, {Bell}, {Bell},
  {Bentum}, {Bernardi}, {Best}, {Breitling}, {Broderick}, {Butcher}, {Ciardi},
  {Dettmar}, {Eisloeffel}, {Frieswijk}, {Gankema}, {Garrett}, {Gerbers},
  {Griessmeier}, {Gunst}, {Hassall}, {Hessels}, {Hoeft}, {Horneffer},
  {Karastergiou}, {K{\"o}hler}, {Koopman}, {Kuniyoshi}, {Kuper}, {Maat},
  {Mann}, {Mevius}, {Mulcahy}, {Munk}, {Nijboer}, {Noordam}, {Paas}, {Pandey},
  {Pandey}, {Polatidis}, {Reich}, {Schoenmakers}, {Sluman}, {Smirnov}, {Sobey},
  {Stappers}, {Swinbank}, {Tagger}, {Tang}, {van Bemmel}, {van Cappellen}, {van
  Duin}, {van Haarlem}, {van Leeuwen}, {Vermeulen}, {Vocks}, {White}, {Wise},
  {Wucknitz}, \& {Zarka}}]{deGasperin_etal:12}
{de Gasperin}, F., {Orr{\'u}}, E., {Murgia}, M., {et~al.} 2012, \aap, 547, A56

\bibitem[{{Doeleman}(2010)}]{2010evn..confE..53D}
{Doeleman}, S. 2010, in 10th European VLBI Network Symposium and EVN Users
  Meeting: VLBI and the New Generation of Radio Arrays, 53

\bibitem[{{Doeleman} {et~al.}(2009){Doeleman}, {Agol}, {Backer}, {Baganoff},
  {Bower}, {Broderick}, {Fabian}, {Fish}, {Gammie}, {Ho}, {Honman},
  {Krichbaum}, {Loeb}, {Marrone}, {Reid}, {Rogers}, {Shapiro}, {Strittmatter},
  {Tilanus}, {Weintroub}, {Whitney}, {Wright}, \&
  {Ziurys}}]{2009astro2010S..68D}
{Doeleman}, S., {Agol}, E., {Backer}, D., {et~al.} 2009, in ArXiv Astrophysics
  e-prints, Vol. 2010, astro2010: The Astronomy and Astrophysics Decadal
  Survey, 68

\bibitem[{{Doeleman} {et~al.}(2008)}]{Doel_etal:08}
{Doeleman}, S.~S., {et~al.} 2008, \nat, 455, 78

\bibitem[{{Doeleman} {et~al.}(2012){Doeleman}, {Fish}, {Schenck}, {Beaudoin},
  {Blundell}, {Bower}, {Broderick}, {Chamberlin}, {Freund}, {Friberg},
  {Gurwell}, {Ho}, {Honma}, {Inoue}, {Krichbaum}, {Lamb}, {Loeb}, {Lonsdale},
  {Marrone}, {Moran}, {Oyama}, {Plambeck}, {Primiani}, {Rogers}, {Smythe},
  {SooHoo}, {Strittmatter}, {Tilanus}, {Titus}, {Weintroub}, {Wright}, {Young},
  \& {Ziurys}}]{Doel_etal:12}
{Doeleman}, S.~S., {Fish}, V.~L., {Schenck}, D.~E., {et~al.} 2012, Science,
  338, 355

\bibitem[{{Done} \& {Gierli{\'n}ski}(2003)}]{Done_Gierlinski_2003}
{Done}, C., \& {Gierli{\'n}ski}, M. 2003, \mnras, 342, 1041

\bibitem[{{Fish} {et~al.}(2011)}]{Fish_etal:11}
{Fish}, V.~L., {et~al.} 2011, \apjl, 727, L36

\bibitem[{{Forman} {et~al.}(2005){Forman}, {Nulsen}, {Heinz}, {Owen}, {Eilek},
  {Vikhlinin}, {Markevitch}, {Kraft}, {Churazov}, \& {Jones}}]{Forman_etal:05}
{Forman}, W., {Nulsen}, P., {Heinz}, S., {et~al.} 2005, \apj, 635, 894

\bibitem[{{Forman} {et~al.}(2007){Forman}, {Jones}, {Churazov}, {Markevitch},
  {Nulsen}, {Vikhlinin}, {Begelman}, {B{\"o}hringer}, {Eilek}, {Heinz},
  {Kraft}, {Owen}, \& {Pahre}}]{Forman_etal:07}
{Forman}, W., {Jones}, C., {Churazov}, E., {et~al.} 2007, \apj, 665, 1057

\bibitem[{{Garcia} {et~al.}(2001){Garcia}, {McClintock}, {Narayan}, {Callanan},
  {Barret}, \& {Murray}}]{G01}
{Garcia}, M.~R., {McClintock}, J.~E., {Narayan}, R., {et~al.} 2001, \apjl, 553,
  L47

\bibitem[{{Gebhardt} {et~al.}(2011){Gebhardt}, {Adams}, {Richstone}, {Lauer},
  {Faber}, {G{\"u}ltekin}, {Murphy}, \& {Tremaine}}]{2011ApJ...729..119G}
{Gebhardt}, K., {Adams}, J., {Richstone}, D., {et~al.} 2011, \apj, 729, 119

\bibitem[{{Ghosh} \& {Abramowicz}(1997)}]{GA97}
{Ghosh}, P., \& {Abramowicz}, M.~A. 1997, \mnras, 292, 887

\bibitem[{{Giroletti} {et~al.}(2012){Giroletti}, {Hada}, {Giovannini},
  {Casadio}, {Beilicke}, {Cesarini}, {Cheung}, {Doi}, {Krawczynski}, {Kino},
  {Lee}, \& {Nagai}}]{Giroletti_etal:12}
{Giroletti}, M., {Hada}, K., {Giovannini}, G., {et~al.} 2012, \aap, 538, L10

\bibitem[{{Hada} {et~al.}(2011){Hada}, {Doi}, {Kino}, {Nagai}, {Hagiwara}, \&
  {Kawaguchi}}]{2011Natur.477..185H}
{Hada}, K., {Doi}, A., {Kino}, M., {et~al.} 2011, \nat, 477, 185

\bibitem[{{Hada} {et~al.}(2013){Hada}, {Kino}, {Doi}, {Nagai}, {Honma},
  {Hagiwara}, {Giroletti}, {Giovannini}, \& {Kawaguchi}}]{2013ApJ...775...70H}
{Hada}, K., {Kino}, M., {Doi}, A., {et~al.} 2013, \apj, 775, 70

\bibitem[{{Heinz} \& {Begelman}(1997)}]{Hein-Bege:97}
{Heinz}, S., \& {Begelman}, M.~C. 1997, \apj, 490, 653

\bibitem[{{Igumenshchev} {et~al.}(2003){Igumenshchev}, {Narayan}, \&
  {Abramowicz}}]{2003ApJ...592.1042I}
{Igumenshchev}, I.~V., {Narayan}, R., \& {Abramowicz}, M.~A. 2003, \apj, 592,
  1042

\bibitem[{{Illarionov} \& {Sunyaev}(1975)}]{Illarionov_Sunyaev_1975}
{Illarionov}, A.~F., \& {Sunyaev}, R.~A. 1975, \aap, 39, 185

\bibitem[{{Kormendy} \& {Ho}(2013)}]{KH13}
{Kormendy}, J., \& {Ho}, L.~C. 2013, \araa, 51, 511

\bibitem[{{Kulkarni} \& {Romanova}(2008)}]{Kulkarni_Romanova_2008}
{Kulkarni}, A.~K., \& {Romanova}, M.~M. 2008, \mnras, 386, 673

\bibitem[{{Lauer} {et~al.}(1992){Lauer}, {Faber}, {Lynds}, {Baum}, {Ewald},
  {Groth}, {Hester}, {Holtzman}, {Kristian}, {Light}, {O'Neil}, {Schneider},
  {Shaya}, \& {Westphal}}]{Laue_etal:92}
{Lauer}, T.~R., {Faber}, S.~M., {Lynds}, R.~C., {et~al.} 1992, \aj, 103, 703

\bibitem[{{Lauer} {et~al.}(1995){Lauer}, {Ajhar}, {Byun}, {Dressler}, {Faber},
  {Grillmair}, {Kormendy}, {Richstone}, \& {Tremaine}}]{Laue_etal:95}
{Lauer}, T.~R., {Ajhar}, E.~A., {Byun}, Y.-I., {et~al.} 1995, \aj, 110, 2622

\bibitem[{{Li} \& {Narayan}(2004)}]{LN04}
{Li}, L.-X., \& {Narayan}, R. 2004, \apj, 601, 414

\bibitem[{{Livio} {et~al.}(1999){Livio}, {Ogilvie}, \& {Pringle}}]{LOP99}
{Livio}, M., {Ogilvie}, G.~I., \& {Pringle}, J.~E. 1999, \apj, 512, 100

\bibitem[{{Madrid}(2011)}]{Madr:PC}
{Madrid}, J. 2011, Private Communication

\bibitem[{{Madrid}(2009)}]{Madr:09}
{Madrid}, J.~P. 2009, \aj, 137, 3864

\bibitem[{{Maoz} {et~al.}(2005){Maoz}, {Nagar}, {Falcke}, \&
  {Wilson}}]{Maoz_etal:05}
{Maoz}, D., {Nagar}, N.~M., {Falcke}, H., \& {Wilson}, A.~S. 2005, \apj, 625,
  699

\bibitem[{{Mathur}(2011)}]{2011CQGra..28l5010M}
{Mathur}, S.~D. 2011, Classical and Quantum Gravity, 28, 125010

\bibitem[{{Mathur}(2014)}]{2014arXiv1406.0807M}
---. 2014, ArXiv e-prints, arXiv:1406.0807

\bibitem[{{Mazur} \& {Mottola}(2004)}]{Mazur_Mottola_2004}
{Mazur}, P.~O., \& {Mottola}, E. 2004, Proceedings of the National Academy of
  Science, 101, 9545

\bibitem[{{McKinney} {et~al.}(2012){McKinney}, {Tchekhovskoy}, \&
  {Blandford}}]{MTB12}
{McKinney}, J.~C., {Tchekhovskoy}, A., \& {Blandford}, R.~D. 2012, \mnras, 423,
  3083

\bibitem[{{Meyer} {et~al.}(2013){Meyer}, {Sparks}, {Biretta}, {Anderson},
  {Sohn}, {van der Marel}, {Norman}, \& {Nakamura}}]{Meyer_etal:13}
{Meyer}, E.~T., {Sparks}, W.~B., {Biretta}, J.~A., {et~al.} 2013, \apjl, 774,
  L21

\bibitem[{{Mills}(1952)}]{M52}
{Mills}, B.~Y. 1952, \nat, 170, 1063

\bibitem[{{Nakamura} \& {Asada}(2013)}]{2013ApJ...775..118N}
{Nakamura}, M., \& {Asada}, K. 2013, \apj, 775, 118

\bibitem[{{Narayan} {et~al.}(1997){Narayan}, {Garcia}, \& {McClintock}}]{N97}
{Narayan}, R., {Garcia}, M.~R., \& {McClintock}, J.~E. 1997, \apjl, 478, L79

\bibitem[{{Narayan} \& {Heyl}(2002)}]{NH02}
{Narayan}, R., \& {Heyl}, J.~S. 2002, in American Institute of Physics
  Conference Series, Vol. 624, Cosmology and Elementary Particle Physics, ed.
  B.~N. {Kursunoglu}, S.~L. {Mintz}, \& A.~{Perlmutter}, 122--131

\bibitem[{{Narayan} {et~al.}(2003){Narayan}, {Igumenshchev}, \&
  {Abramowicz}}]{NIA03}
{Narayan}, R., {Igumenshchev}, I.~V., \& {Abramowicz}, M.~A. 2003, \pasj, 55,
  L69

\bibitem[{{Narayan} {et~al.}(1998){Narayan}, {Mahadevan}, {Grindlay}, {Popham},
  \& {Gammie}}]{Nara_etal:98}
{Narayan}, R., {Mahadevan}, R., {Grindlay}, J.~E., {Popham}, R.~G., \&
  {Gammie}, C. 1998, \apj, 492, 554

\bibitem[{{Narayan} \& {McClintock}(2008)}]{NM08}
{Narayan}, R., \& {McClintock}, J.~E. 2008, New Astronomy Reviews, 51, 733

\bibitem[{{Narayan} {et~al.}(2012){Narayan}, {S{\c a}dowski}, {Penna}, \&
  {Kulkarni}}]{NSPK12}
{Narayan}, R., {S{\c a}dowski}, A., {Penna}, R.~F., \& {Kulkarni}, A.~K. 2012,
  \mnras, 426, 3241

\bibitem[{{Novikov} \& {Thorne}(1973)}]{NT73}
{Novikov}, I.~D., \& {Thorne}, K.~S. 1973, in Black Holes (Les Astres Occlus),
  ed. C.~{Dewitt} \& B.~S. {Dewitt}, 343--450

\bibitem[{{Owen} {et~al.}(2000){Owen}, {Eilek}, \& {Kassim}}]{OEK00}
{Owen}, F.~N., {Eilek}, J.~A., \& {Kassim}, N.~E. 2000, \apj, 543, 611

\bibitem[{{Perlman} {et~al.}(2001){Perlman}, {Biretta}, {Sparks}, {Macchetto},
  \& {Leahy}}]{Perl_etal:01a}
{Perlman}, E.~S., {Biretta}, J.~A., {Sparks}, W.~B., {Macchetto}, F.~D., \&
  {Leahy}, J.~P. 2001, \apj, 551, 206

\bibitem[{{Perlman} {et~al.}(2011){Perlman}, {Adams}, {Cara}, {Bourque},
  {Harris}, {Madrid}, {Simons}, {Clausen-Brown}, {Cheung}, {Stawarz},
  {Georganopoulos}, {Sparks}, \& {Biretta}}]{Perl_etal:11}
{Perlman}, E.~S., {Adams}, S.~C., {Cara}, M., {et~al.} 2011, \apj, 743, 119

\bibitem[{{Rafferty} {et~al.}(2006){Rafferty}, {McNamara}, {Nulsen}, \&
  {Wise}}]{Rafferty_etal:06}
{Rafferty}, D.~A., {McNamara}, B.~R., {Nulsen}, P.~E.~J., \& {Wise}, M.~W.
  2006, \apj, 652, 216

\bibitem[{{Rees}(1984)}]{1984ARA&A..22..471R}
{Rees}, M.~J. 1984, \araa, 22, 471

\bibitem[{{Reynolds} {et~al.}(1996){Reynolds}, {Fabian}, {Celotti}, \&
  {Rees}}]{RFCR96}
{Reynolds}, C.~S., {Fabian}, A.~C., {Celotti}, A., \& {Rees}, M.~J. 1996,
  \mnras, 283, 873

\bibitem[{{S{\c a}dowski} {et~al.}(2013){S{\c a}dowski}, {Narayan}, {Penna}, \&
  {Zhu}}]{SNPZ13}
{S{\c a}dowski}, A., {Narayan}, R., {Penna}, R., \& {Zhu}, Y. 2013, \mnras,
  436, 3856

\bibitem[{{S\'ersic}(1968)}]{Sers:68}
{S\'ersic}, J.~L. 1968, {Atlas de galaxias australes}

\bibitem[{{Sparks} {et~al.}(1996){Sparks}, {Biretta}, \&
  {Macchetto}}]{Spar-Bire-Macc:96}
{Sparks}, W.~B., {Biretta}, J.~A., \& {Macchetto}, F. 1996, \apj, 473, 254

\bibitem[{{Spruit} {et~al.}(1995){Spruit}, {Stehle}, \&
  {Papaloizou}}]{Spruit_et_al_1995}
{Spruit}, H.~C., {Stehle}, R., \& {Papaloizou}, J.~C.~B. 1995, \mnras, 275,
  1223

\bibitem[{{Stawarz} {et~al.}(2006){Stawarz}, {Aharonian}, {Kataoka},
  {Ostrowski}, {Siemiginowska}, \& {Sikora}}]{Staw_etal:06}
{Stawarz}, {\L}., {Aharonian}, F., {Kataoka}, J., {et~al.} 2006, \mnras, 370,
  981

\bibitem[{{Tchekhovskoy}(2015)}]{2015ASSL..414...45T}
{Tchekhovskoy}, A. 2015, in Astrophysics and Space Science Library, Vol. 414,
  Astrophysics and Space Science Library, ed. I.~{Contopoulos}, D.~{Gabuzda},
  \& N.~{Kylafis}, 45

\bibitem[{{Tchekhovskoy} {et~al.}(2010){Tchekhovskoy}, {Narayan}, \&
  {McKinney}}]{TNM10}
{Tchekhovskoy}, A., {Narayan}, R., \& {McKinney}, J.~C. 2010, \apj, 711, 50

\bibitem[{{Tchekhovskoy} {et~al.}(2011){Tchekhovskoy}, {Narayan}, \&
  {McKinney}}]{TNM11}
---. 2011, \mnras, 418, L79

\bibitem[{{Walsh} {et~al.}(2013){Walsh}, {Barth}, {Ho}, \&
  {Sarzi}}]{2013ApJ...770...86W}
{Walsh}, J.~L., {Barth}, A.~J., {Ho}, L.~C., \& {Sarzi}, M. 2013, \apj, 770, 86

\bibitem[{{Young} {et~al.}(2002){Young}, {Wilson}, \& {Mundell}}]{YWM02}
{Young}, A.~J., {Wilson}, A.~S., \& {Mundell}, C.~G. 2002, \apj, 579, 560

\bibitem[{{Young} {et~al.}(1978){Young}, {Westphal}, {Kristian}, {Wilson}, \&
  {Landauer}}]{Youn_etal:78}
{Young}, P.~J., {Westphal}, J.~A., {Kristian}, J., {Wilson}, C.~P., \&
  {Landauer}, F.~P. 1978, \apj, 221, 721

\bibitem[{{Yuan} \& {Narayan}(2014)}]{YN14}
{Yuan}, F., \& {Narayan}, R. 2014, \araa, 52, 529

\end{thebibliography}
\bibliographystyle{apj}

\end{document}